\documentclass{egpubl}
\usepackage{eurovis2022}

\SpecialIssuePaper         

\usepackage[T1]{fontenc}
\usepackage{dfadobe}
\usepackage{xspace}

\usepackage{cite}  
\BibtexOrBiblatex
\electronicVersion
\PrintedOrElectronic
\ifpdf \usepackage[pdftex]{graphicx} \pdfcompresslevel=9
\else \usepackage[dvips]{graphicx} \fi

\usepackage{egweblnk}

\usepackage{mfirstuc}

\usepackage{titlesec}

\titlespacing*{\section}{0pt}{1pt plus 1pt minus 1pt}{0pt plus 1pt minus 1pt}
\titlespacing*{\subsection}{0pt}{1pt plus 1pt minus 1pt}{0pt plus 1pt minus 1pt}
\titlespacing*{\subsubsection}{0pt}{1pt plus 1pt minus 1pt}{0pt plus 1pt minus 1pt}
\newcommand{\squeezeafterfigure}{\vspace{-0.31in}}

\newcommand{\ie}{\textit{i.e.}}
\newcommand{\etal}{\textit{et al.}}
\MFUnocap{in}
\MFUnocap{of}
\MFUnocap{on}
\MFUnocap{with}
\MFUnocap{and}
\MFUnocap{at}
\newcommand{\tag}[1]{\textit{\capitalisewords{#1}}}
\newcommand{\cat}[1]{\textbf{\capitalisewords{#1}}}
\newcommand{\rdataisugly}{\textit{r/dataisugly}\xspace}

\usepackage[normalem]{ulem} 
\usepackage{xcolor}

\ifx\revision\undefined
    \newcommand{\leo}[2]{{#2}}
    \newcommand{\leobf}[2]{{#2}}
    \newcommand{\leonotice}[1]{}
\else
    \newcommand{\leo}[2]{{\sout{#1}\color{orange}{#2}}}
    \newcommand{\leobf}[2]{{\sout{#1}\textbf{#2}}}
    \newcommand{\leonotice}[1]{\leo{}{#1}}
\fi

\title[Misinformed by Visualization]%
      {Misinformed by Visualization: What Do We Learn From Misinformative Visualizations?}

\author[L. Y. Lo \& A. Gupta \& K. Shigyo \& A. Wu \& E. Bertini \& H. Qu]
{
    \parbox{\textwidth}{\centering
        Leo Yu-Ho Lo$^{1}$\orcid{0000-0002-3660-3765}, 
        Ayush Gupta$^{1}$\orcid{0000-0002-3782-3786},
        Kento Shigyo$^{1}$\orcid{0000-0002-5095-7500},
        Aoyu Wu$^{1}$\orcid{0000-0001-9187-9265},
        Enrico Bertini$^{2}$\orcid{0000-0002-9932-0551},
        and Huamin Qu$^{1}$\orcid{0000-0002-3344-9694}
    }
    \\
    \parbox{\textwidth}{\centering
        $^1$Hong Kong University of Science and Technology \\
        $^2$Northeastern University
    }
}

\begin{document}


\maketitle
\begin{abstract}
Data visualization is powerful in persuading an audience.
However, when it is done poorly or maliciously, a visualization may become misleading or even deceiving.
Visualizations give further strength to the dissemination of misinformation on the Internet.
The visualization research community has long been aware of visualizations that misinform the audience, mostly associated with the terms ``lie'' and ``deceptive.''
Still, these discussions have focused only on a handful of cases.
To better understand the landscape of misleading visualizations, we open-coded over one thousand real-world visualizations that have been reported as misleading.
From these examples, we discovered 74 types of issues and formed a taxonomy of misleading elements in visualizations.
We found four directions that the research community can follow to widen the discussion on misleading visualizations: (1) informal fallacies in visualizations, (2) exploiting conventions and data literacy,  (3) deceptive tricks in uncommon charts, and (4) understanding the designers' dilemma.
This work lays the groundwork for these research directions, especially in understanding, detecting, and preventing them.


\vspace{10pt}

\ccsdesc[500]{Human-centered computing~Information visualization}

\printccsdesc   
\end{abstract}


\section{Introduction}
\label{intro}

\leonotice{Revision changes are highlighted in orange.}

Data visualization has become a part of our daily information consumption.
The general public consumes information from visualizations appearing in newspapers, on television, and on the Internet.
Some of these visualizations have the purpose of swaying the audience toward a particular agenda.
For example, they are used in political promotions, business advertisements, donation programs, and even on electricity bills to persuade us to reduce our energy consumption \cite{thaler2008nudge}.
The study by Pandey \etal~showed that data visualization has a significant effect on fortifying a message or prompting a change \cite{pandey2014persuasive}.
When visualizing the supporting data faithfully, these usages are good demonstrations of the power of data visualization.
However, when the data does not support the intended claims, it is tempting to distort the visualizations to make them look supportive, which is misleading.

The existence of misleading visualization cases has long been documented and discussed.
Back in the 1950s, before personal computers, the book \textit{How to Lie with Statistics} by Huff \cite{huff1993lie} collected examples of misleading charts from the newspapers.
Nowadays, we have a much larger data source on the Internet.
Still, the discussions among the research community are still focused on a limited set of cases -- most commonly, truncated axis, area encoding, and 3D charts \cite{pandey2015deceptive, cairo2015graphics, correll2017black, mcnutt2018linting, szafir2018good, moritz2018formalizing, zhengvisualization, hopkins2020visualint, lauer2020deceptive}.
We imagine that taking a broad look at the problem will help further the discussion, and perhaps we may find commonalities in other cases that may lead to breakthroughs on these long-discussed cases.

\begin{figure}
  \centering
  \includegraphics[width=\linewidth, height=0.94\textheight, keepaspectratio, clip, trim=0cm 0cm 10.36cm 0cm]{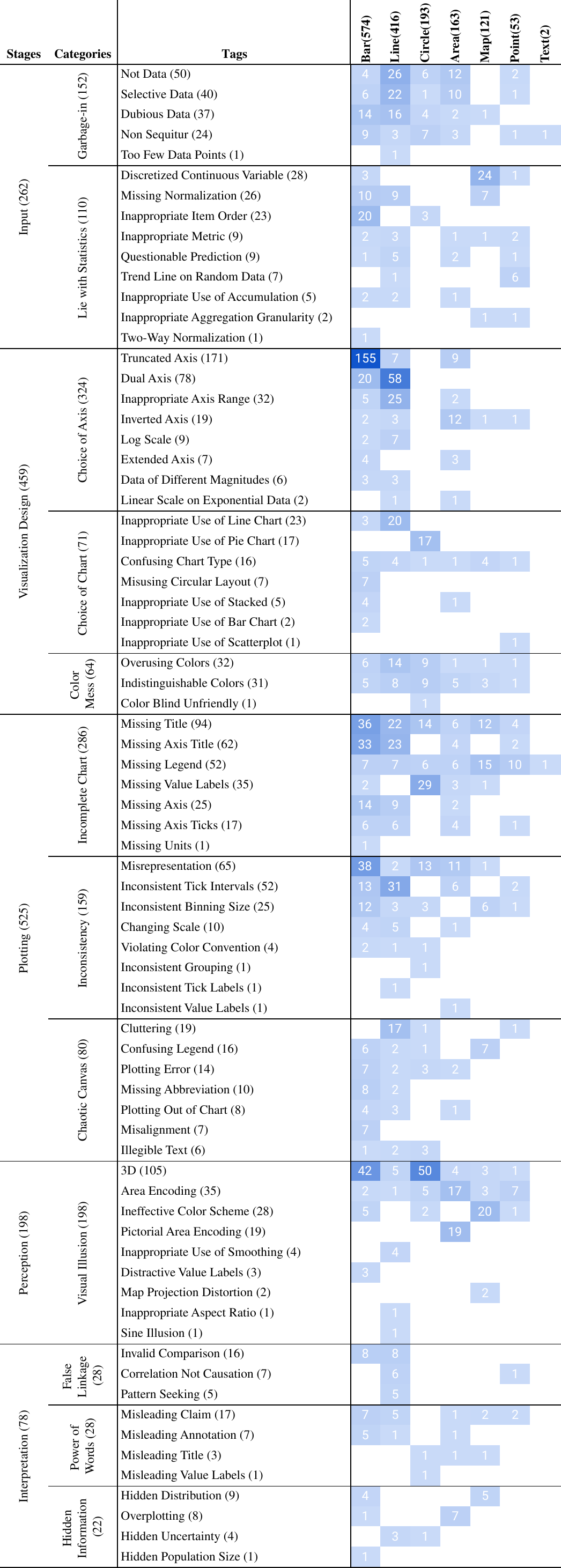}
  \caption{Taxonomy of the 74 issues discovered in the 12 categories, which are further grouped into the five analytics stages. The tag descriptions can be found in the Appendix.}
  \label{fig:taxonomy}
\end{figure}

To widen our horizon on misleading visualization, we have collected 129,125 visualization images from the Internet through keyword searches from three major search engines and four social media platforms.
We open-coded the collected images until we reached saturation, \ie, after a period of no discovery of any new type of issues.
At the end of the coding process, we had examined 6,500 images (\char`\~5\% of the collected images), and among these images, 1,979 images (\char`\~30\% of examined images) were candidate images--visualizations that are the target of this study--and 1,143 (\char`\~58\% of candidate images) were tagged with issues.
We discovered 74 different types of issues and organized them into a taxonomy of 12 categories.
These 12 categories further fit into a five-stage visualization process.
\autoref{fig:taxonomy} shows the 74 types of issues in a five-stage visualization process.

From the taxonomy and the notes taken during the coding process, we identified four directions that would enrich the discussion on misleading visualizations: (1) informal fallacies in visualizations, (2) exploiting conventions and data literacy, (3) deceptive tricks in uncommon charts, and (4) providing design guidelines by understanding the designers’ dilemma.

Informal fallacy sneaks in at the input and interpretation stages of the visualization process.
For example, cherry-picking or double counting the data and false analogies or causal linkages in the message interpreted from the visualization.
We suspect that most of the informal fallacies in argumentation also apply to visualizations.
Secondly, miscommunication happens when the convention is broken implicitly, and it happens with visualizations, too.
There are many implicit expectations when viewing a chart, for example, expecting a linear scale but the data is shown in log scale, and expecting consistent intervals between tick marks but the scale is changing along the axis.
Charts become misleading when the implicit assumptions are not met.
Thirdly, a word cloud is a common visualization form, but it is rarely reported as misleading.
Are word clouds immune to being misleading?
Or, are misleading word clouds stealthy enough to go unspotted?
Are there any other visualization forms that share this commonality?
These are the questions we are eager to answer, and they may lead us in a new research direction.
Lastly, not all misleading visualizations intend to mislead.
We have seen examples where the visualization authors have made their best design effort to avoid misleading the audience, given the limitation of the underlying data and presentation form.
For example, the age groups in the dataset are unevenly split, so avoiding an inconsistent binning size is challenging, or a projection of the spherical Earth onto a 2D world map is unavoidably distorted.
Understanding these dilemmas in misleading visualizations is an important step toward providing practical guidelines for visualization practitioners.

\section{Related Work}
\label{related}

\subsection{Misinformation}

The Internet is filled with false information in the post-truth era \cite{keyes2004post}.
The current study of misinformation has been focused on the textual content and propagation network, but the equally important visual content is being neglected \cite{wardle2017information}.
The false information that has been circulated online during the pandemic is an unfortunate demonstration of the key role played by visual content.
``Counter-visualizations,'' a term created by Lee \etal~\cite{lee2021viral} to describe the visualizations used in unorthodox ways, are widely circulated on social media platforms to challenge public health measures.
The COVID-19-related counter-visualizations and their associated discussions on social media platforms show that the use of a powerful persuasive device--data visualization--can work against scientific reasoning.
We echo the authors’ call that it is a matter of urgency for the research community to pay more attention to the visualizations created by non-experts and those circulated online.

\subsection{Lie with Visualizations}

Studies on misleading visualizations began a long time ago, before the Internet.
Back in the 1950s, Huff’s classic statistics book, \textit{How to Lie with Statistics}, included misleading visualizations collected from newspapers, illustrating how journalists applied tricks to distort the actual meaning of the underlying data \cite{huff1993lie}.
He showed that the problem of misleading visualizations is highly related to the nature of statistical communication.
This was at a time when personal computers had not yet been invented.

In the age of computer-drawn visualizations, the discussion continued with wider coverage by Tufte in the 1980s, among others.
In his book \textit{The Visual Display of Quantitative Information}, Tufte introduced the terms ``graphical integrity'' and ``lie factor,'' stating that visual encoding should proportionally represent the numerically measured values in a visualization \cite{tufte1983visual}.
Despite the rule's simplicity, it captures many basic cases of misleading visualizations--for example, truncated axis, area encoding, and 3D pie charts.
However, the rule's controversy is related to its over-simplicity on the cases not covered, like an inverted axis and cases where violating the rule to present the data visually is unavoidable, such as the valid use of log scale and 2D map projection.

In a recent book by Cairo entitled \textit{How Charts Lie}, the discussion is more focused on spotting the dubious data underlying the visualizations and the intended message of the visualizations' authors \cite{cairo2019charts}.
Using the examples collected from current affairs and the Internet, the author revealed how visualization creators could hide the underlying \leo{questionable}{dubious} data and present their intended message to the audience.
Instead of misleading the audience by manipulating the visual elements that can be spotted from the visualizations themselves, like truncated axis, or 3D charts, manipulating the data is much harder to spot.
As the author quoted Ronald Coase, ``If you torture data long enough, it’ll always confess to anything.''
These examples also reflect the reality of the digital world we live in.
The abundance of content keeps us busy, so we have no time to verify it.
This type of misleading content is impossible to identify from the visualization image alone without understanding the context.

Monmonier’s \textit{How to Lie with Maps} \cite{monmonier2018lie} follows the spirit of the 1950s classic and focuses on how cartographic visualizations may provide misleading information.
Maps are one of the most common visualization types, and we are familiar with the distorted two-dimensional projection of the three-dimensional spherical Earth.
The commonly used Mercator projection has long resulted in a misunderstanding of Africa's size, which is largely understated on the projected 2D world map.
This case reminds us that the creators are not necessarily ill-intended when making visualizations that might eventually mislead people.
Instead, it is a design trade-off in a dilemma situation.
Technological advancements may provide solutions to these dilemmas.
For example, Google Maps has been designed to show the actual spherical size on a 2D screen.
We encountered more examples of these design dilemmas, and they are not limited to maps.
They are in various visual forms and have different rationales inherited from the underlying data that cause these design dilemmas.

The above books are good collections of misleading visualization examples.
This work aims to complement and expand their work to provide broader coverage of misleading visualizations on the Internet.
From our collected visualization examples, we found that some misleading visualizations indeed try to add visual cues to avoid misleading the audience.
Despite the best efforts of the designers, it is not enough to prevent the audience from getting the wrong message.
Understanding these dilemmas will be a step forward for the community in developing practical guidelines for designers to make better design decisions.

\subsection{Academic Research on Misleading Visualizations}

The motivation of this work is to open up more research directions for the study of misleading visualizations and look into the latest developments in misleading tricks being used in visualizations.
Several previous works have summarized from the literature the potential pitfalls when creating visualizations \cite{bresciani2015pitfalls, mcnutt2020surfacing}, optical illusions \cite{bach2006optical}, and cognitive biases \cite{dimara2018task}.
Each of these works includes a rich set of potential causes of misleading visualizations.
Still, the study of misleading visualizations has focused on a few tricks inspired by real-world examples \cite{correll2017black}.

Among many misleading tricks, truncated axis, inverted axis, area encoding, dual-axis, and rainbow colors are the most commonly picked for the studies, usually by heuristics or from the authors’ experiences \cite{pandey2015deceptive, cairo2015graphics, correll2017black, mcnutt2018linting, szafir2018good, moritz2018formalizing, zhengvisualization, hopkins2020visualint, lauer2020deceptive}.
The notorious inverted axis visualization titled ``Gun deaths in Florida'' has been frequently discussed.
In contrast, in our labeled set of real-world examples, inverted axis is a comparatively rare occurrence.
Why is that the case?
On the other hand, are there tricks that we have overlooked but widely used?
This work aims to better understand the domain of misleading visualizations from real-world examples and guide our future research on developing detection and prevention measures \cite{hopkins2020visualint}.

\section{Method}
\label{method}

To build up an understanding of the domain of misleading visualizations, we followed the grounded theory method (GTM) for human-computer interaction studies described by Muller \cite{muller2014curiosity}.
GTM is a widely used method in HCI studies for understanding unchartered domains.
It is implemented through continuous data collection, coding the data, and revising the developing theory until theoretical saturation is reached, also known as ``no further surprises.''

\subsection{Data Collection}

GTM starts with data collection, but collecting a large set of real-world examples of misleading visualizations was challenging.
Fortunate or not, compared to the sea of faithful visualizations, the existence of misleading visualizations is relatively sparse.
For the aforementioned book authors, collection relies on spotting misleading visualizations encountered in daily life, on the Internet, or in the news media.
Such a collection takes a long time and substantial devoted effort, and in the end, the collected dataset may not have representative coverage because of personal bias in the collection process.


We turned our attention to the reported cases on the Internet through web search engines and social media platforms.
There are three major search engines, (1) Google, (2) Bing, and (3) DuckDuckGo, and four major social media platforms, (1) Facebook, (2) Twitter, (3) Pinterest, and (4) Reddit.
For the keyword terms, we combined the ten adjectives or verbs derived from the synonyms of ``misleading'' and five nouns derived from the synonyms of ``visualization'' to form 50 keyword terms.
We applied these keyword searches on search engines and social media platforms.
The exception to this was Reddit, which has a subforum named \textit{r/dataisugly} dedicated to discussing misleading or poorly made visualizations.
We found that the discussion posts on this subforum were more relevant than the results of a keyword search.
Therefore, instead of using a keyword search on Reddit, we crawled all the discussion threads on \rdataisugly.
By crawling the keyword search results from three search engines, three social media platforms, and the \rdataisugly on 24th February 2021, a total number of 129,125 images, including irrelevant images, were collected, and the composition is shown in \autoref{table:keywords}.
While this method provides representative coverage, it is limited to the reported cases, and the unspotted cases are uncovered.
We suspect that there are many more stealthy cases not being spotted.
It remains an important question to explore in future work: what are the characteristics of stealthy misleading visualizations?
We have included further discussions of this question, summarized from our findings, in \autoref{discussion}.

\subsection{Open-Coding}

\begin{table*}
  \caption{Numbers of collected images by data source and keywords. Relevancy\leobf{}{--the number of candidate images in 100 return results--}reflects the relevant images after removing the ``junk'' returned by data sources. Accuracy\leobf{}{--the number of tagged images in 100 return results--}reflects the misinformative or uninformative visualizations returned by data sources. \rdataisugly is a subforum of Reddit dedicated to the discussion of misleading or poorly made visualizations. The highest values are highlighted in \textbf{bold}.}
  \centering
  \includegraphics[width=\textwidth, clip, keepaspectratio, trim=0cm 25.6cm 0cm 0cm]{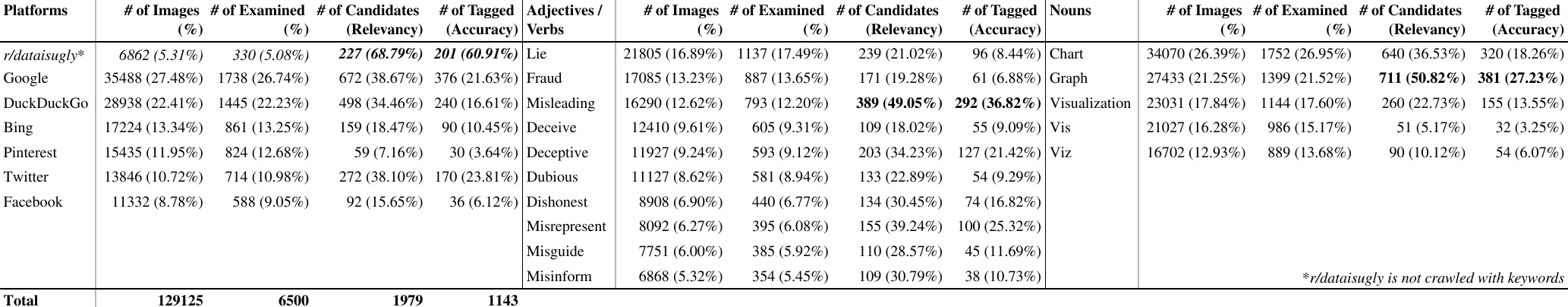}
  \label{table:keywords}
  \squeezeafterfigure
\end{table*}

Since the collected images are mixed with irrelevant images--images that are not visualizations--we filtered the collection first before open-coding.
To clarify the terminology for describing the coding process, we use ``image'' to refer to image files and ``visualization'' to refer to the charts in the image.
Besides relevancy, we further narrowed down the focus of the study to single-viewed visualizations and included only the common chart types, \ie, bar, line, circle, area, map, and point, by following the chart type taxonomy proposed by Borkin \etal~\cite{borkin2013makes}.
Out of curiosity, the previous study by Borkin stated a counter-intuitive fact that text visualizations are rarely seen in their collected dataset.
We wanted to check if this was also the case for our collected images and therefore also included text visualization as our study targets.

To ensure the ordering of the collected images would not affect the coding process, we first performed a random shuffle on the order of all the collected images and then fetched images in batches for filtering and coding.
The open coding process is:
\begin{enumerate}
    \item Fetch a batch of 500 images from the randomly shuffled images.
    \item Filter out irrelevant images.
    \item Filter out visualizations with multiple views.
    \item Filter out visualizations that are not one of the following types: (1) bar, (2) line, (3) circle, (4) area, (5) map, (6) point, or (7) text.
    \item Examine each of the images and the associated social media post and web page to figure out the issues in the visualizations and add one or multiple tags to it.
\end{enumerate}

During the coding process, we found that, although we started with the intention of studying misleading visualizations, the collected visualizations included a lot of uninformative visualizations.
This type of visualization does not mislead, but the reader cannot get any information from it either.
When we read a visualization, we expect to get information from it, but uninformative visualizations simply leave readers confused.
For example, a chart with no axis titles nor units is not comprehensible and provides no useful information regardless of whether it is drawn accurately or not (\autoref{fig:stage2issues}e).
We found that the study of misleading visualizations is inseparable from these uninformative visualizations.
They are both failed cases of data visualization.
To avoid assuming the intention of the visualization authors, we use the collective term ``issues'' to refer to both (1) \textit{distortion}, where the reader gets a different message than the information in the data, and (2) \textit{confusion}, where the reader cannot get information from the visualizations at all.
These two types of issues, respectively, result in \textbf{misinformative} and \textbf{uninformative} visualizations.

Two of the authors open-coded the first two batches, a total of 1,000 images, independently by following the procedure mentioned above and then discussing the definition of the tags.
The undecidable cases were put into a discussion with a third author.
Examples of these undecidable cases are whether it is valid to truncate the y-axis of a line chart or visualization that require extra context to be understood that is not included in the associated social media post or web page.
The authors discussed and looked for more context on the Internet to resolve these cases.
In this way, the two coders iteratively discussed and refined the definition of each tag until they reached Cohen’s $\kappa > 0.7$.
Among the first two batches of 1,000 images, 303 were candidate images, 174 were tagged, and 52 tags were discovered.
Following the codebook, one of the coders continued to code images batch by batch.
Upon discovering new tags, the two coders discussed their definitions and checked if they overlapped with any previously discovered tags to ensure the tags were mutually exclusive.
The coding process continued until the saturation criterion was reached, \ie, no new tag was found in the last 100 tagged images.
In total, 6,500 images were examined, 1,979 of them were candidate images, 1,143 were tagged with at least one issue, and 74 tags were discovered.
To ensure correctness, the coders then performed a within-tag consistency check.
The tagged images are publicly accessible on \href{https://leoyuholo.github.io/bad-vis-browser}{https://leoyuholo.github.io/bad-vis-browser}, and the shuffled list of 129,125 images and the open-coding results can be found on OSF: \href{https://osf.io/wghxd}{https://osf.io/wghxd}.


\section{Data Sources, Keywords and Limitations}
\label{stats}




\autoref{table:keywords} shows the number of results returned from different sources.
Since search engines return results from the same pool of content and users on social media platforms tend to cross-post the same content onto different platforms, these data sources are not mutually exclusive.
One of these cases is the notorious inverted axis visualization ``Gun deaths in Florida.''
From the labeled 19 examples of inverted axis, the ``Gun deaths in Florida'' visualization was repeated 11 times.
Such a dramatic case with the reversal effect of an inverted axis has been widely cross-posted across the Internet.


\leo{The relevancy of the returned results--the number of candidate images in 100 return results--reflects the relevant images after removing the ``junk'' returned by the data sources.}{}
\leo{Additionally, the accuracy of the returned results--the number of tagged images in 100 return results--reflects the fruitfulness of the data sources.}{}
\leo{The Reddit subforum \rdataisugly is a highly concentrated discussion forum.
It has both high relevancy and accuracy.}{The Reddit subforum \rdataisugly has both high relevancy and accuracy, reflecting it as a highly concentrated discussion forum on misinformative and uninformative visualizations.}
Since all these return results are reported cases associated with the keywords or the discussion topics, the unreported cases on the low relevancy and accuracy sources may be prevalent but not spotted or discussed.
The study by Lee \etal~\cite{lee2021viral} on both Twitter and Facebook is a reminder of such prevalence that the collected dataset cannot capture.
The keywords ``misleading,'' ``misrepresent,'' and ``deceptive'' have the highest relevancy and accuracy.
Surprisingly, the keyword ``lie'' has low accuracy despite most related books being titled with it.
Many of the examined images are related to financial data and mention ``Charts Don’t Lie'' in the associated social media posts, which may refer to the serial financial advice book authored by Ryan \cite{ryan2014charts}.
For the nouns, visualizations are most commonly referred to as ``chart'' and ``graph,'' while the short forms ``vis'' and ``viz'' are not.
Studying the visualizations circulating on the Internet is an important yet neglected aspect of understanding information disorder \cite{wardle2017information}.
The reported statistics may inform future studies on choosing keywords and selecting data sources.

Despite our best efforts to gather images from the widest available sources, our coverage was still limited by the following factors:

\begin{enumerate}
    \item Although there are visualizations in languages other than English, the majority of the visualizations are in English.
    \item The visualizations are static images. Videos and animated \leo{GIFs}{visualizations} are not covered.
    \item The sampled population from different data sources has overlaps. \leo{}{Same visualization or its variants--with annotations, corrections, or framing--may be posted more than once. For example, we have seen the notorious ``Gun deaths in Florida'' and its variants 11 times during the coding process.}
    \item The population reflects only the reported cases. \leo{}{Issues that are less discussed or less aware are underrepresented because the coding process relies on explicit reporting and the coders' awareness to spot the issues in the visualizations. For example, color blind unfriendly is more ubiquitous than the only reported case in the tagged images.}
\end{enumerate}

In the coding process, we discovered 74 issues in visualizations.
Examples of the 15 most common issues can be found in the Appendix.
Among these common issues, we can locate the frequently discussed issues of \tag{truncated axis} (1st), \tag{3D} (2nd), \tag{dual-axis} (4th), and \tag{area encoding} (12th).
If \tag{area encoding} is combined with \tag{pictorial area encoding}, it ranks at 8th.
Despite that, there are more common issues missing from the discussions.
The following issues can be misleading: \tag{misrepresentation} (5th), \tag{inconsistent tick interval} (7th), \tag{not data} (9th), \tag{selective data} (10th), and \tag{dubious data} (11th).
These issues can also be confusing in that the readers cannot get any information from the visualizations: \tag{missing title} (3rd), \tag{missing axis title} (6th), \tag{missing legend} (8th), \tag{missing value labels} (13th), \tag{inappropriate axis range} (14th), and \tag{overusing colors} (15th).
These issues are not only related to the perceptual understanding of a chart, like \tag{3D} or \tag{area encoding}, but also involve a wider spectrum of causes, like input data, missing chart elements, axis choice, and scalability.
We have discovered 74 issues, but they are not exhaustive to the issue space.
Putting the issues into a taxonomy will help us learn more about the issue space and understand what causes visualizations to be misinformative or uninformative.

\section{Taxonomy}
\label{taxonomy}


Our first intuitive attempt at axial coding was to group issues by their concerned chart elements.
For example, \tag{dual axis} concerns the axis, hence it is grouped with other issues related to the axis, such as \tag{truncated axis}, \tag{missing axis}, and \tag{inconsistent tick intervals}.
Following the same approach, the issues related to data are grouped together.
For example, the \tag{data of different magnitudes} issue is grouped with \tag{selective data}.
These groups are related to (1) chart elements: axis, color, legend, and annotation, (2) grammar of graphics \cite{wickham2010layered}: data, statistical transformation, and aesthetic mapping, and (3) other: plotting and message.
Although the elements of these groups share similarities, they do not help to clarify why a visualization misleads or when the misleading element was introduced.

Our second attempt was guided by the notes taken during the coding process when examining the visualizations and the associated social media posts or web pages.
One of the visualizations examined is associated with a blog post explaining the dual axis that states ``Dual axes time series plots may be ok sometimes after all'' \cite{ellis2016dual}.
The \tag{dual axis} issue is highly related to the issue \tag{data of different magnitudes}, where a \tag{dual axis} is trying to solve the dilemma of plotting two data series of different magnitudes.
Take \autoref{fig:stage2issues}g as an example.
When forcing the two series to be plotted on the same y-axis, the smaller magnitude series gets dominated by the larger one, and the reader can barely see it exists.
Instead of grouping \tag{data of different magnitudes} with other data-related issues, it is associated with other issues concerning the \cat{Choice of Axes}.
This grouping gives us a clearer picture of why a visualization becomes misleading.
Other decisions on the \cat{choice of axes} that resulted in misleading visualizations include \tag{inappropriate axis range}, \tag{log scale}, \tag{linear scale on exponential data}, and the trio \tag{Truncated/Extended/Inverted axis}.
\autoref{fig:stage2issues} shows the example visualizations for this category, and a description of these tags can be found in the Appendix.

\tag{Missing axis}, \tag{missing axis title}, and \tag{missing axis ticks} are related to the axis, but they are not design decisions.
Putting them alongside \tag{truncated axis} or \tag{log scale} would blur the picture rather than clarify it.
Instead, we put all issues related to missing parts of the chart into the category \cat{Incomplete Chart}.
This grouping provides a better foundation to develop theories upon.
We can further fit these categories into a five-stage visual analytics process modified from McNutt \etal~\cite{mcnutt2020surfacing}.
These stages are (1) input stage, (2) visualization design stage, (3) plotting stage, (4) perception stage, and (5) interpretation stage.

\subsection{Input Stage: Data Curation and Wrangling}


\begin{figure*}
  \centering
  \includegraphics[width=\linewidth, height=\textheight, keepaspectratio, clip, trim=0cm 0cm 0cm 0cm]{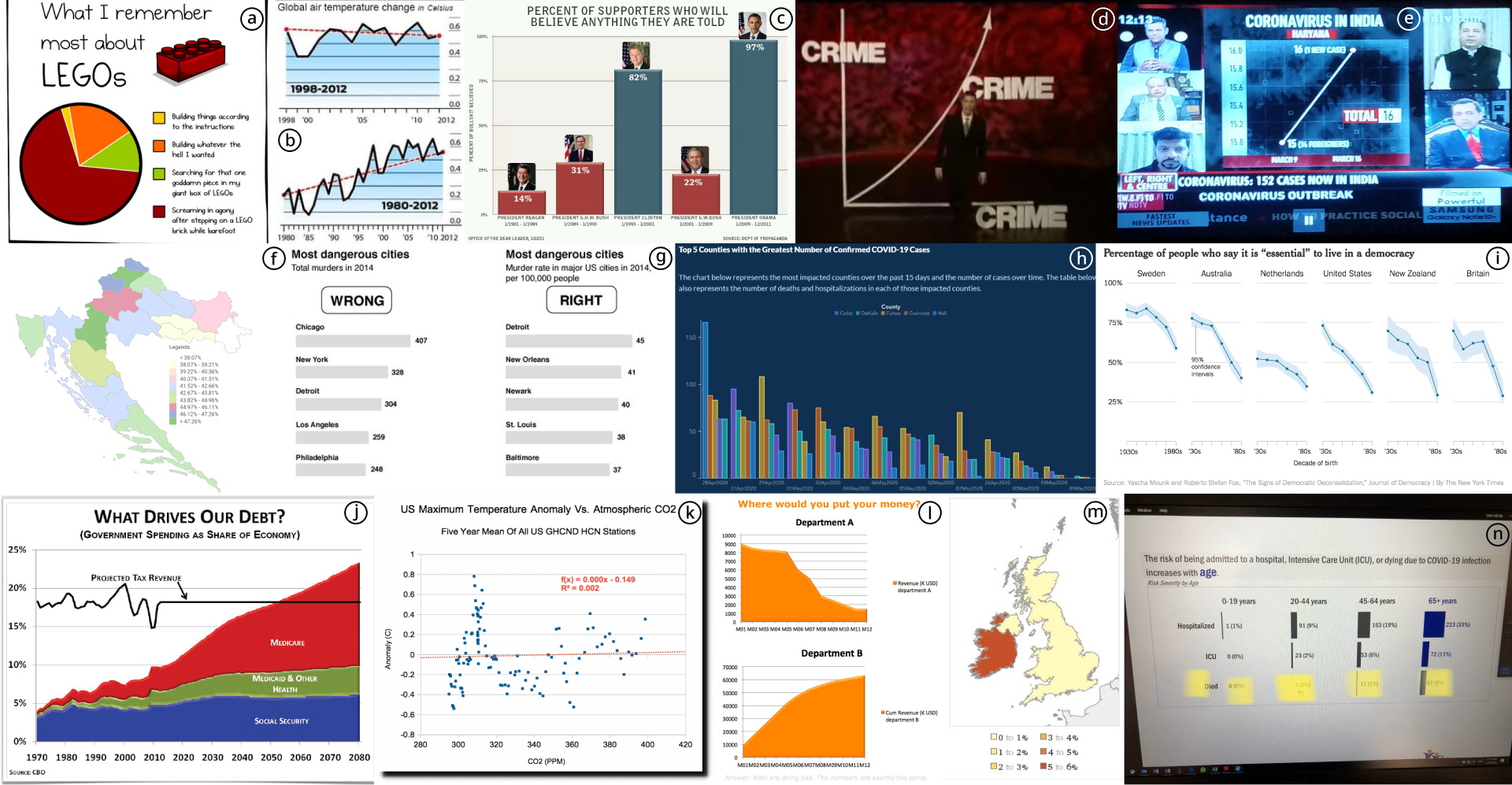}
  \caption{Issues in the Input Stage. These issues are (a) Not Data, (b) Selective Data, (c) Dubious Data, (d) Non Sequitur, (e) Too Few Data Points, (f) Discretized Continuous Variable, (g) Missing Normalization, (h) Inappropriate Item Order, (i) Inappropriate Metric, (j) Questionable Prediction, (k) Trend Line on Random Data, (l) Inappropriate Use of Accumulation, (m) Inappropriate Aggregation Granularity, and (n) Two-way Normalization. See the Appendix for an explanation of each visualization, accompanied by a high-resolution version of the image.}
  \label{fig:stage1issues}
  \squeezeafterfigure
\end{figure*}

The \cat{Garbage-in} category borrows the idiom ``garbage in, garbage out'' to describe the issues related to inputting faulty data into the process, which means outputting a faulty visualization is unavoidable.
\tag{Not Data} are visualizations that are drawn to illustrate concepts rather than visualizing data.
Readers may mistake these as being drawn according to actual data, but this is not the intention.
Similarly, readers expect \tag{Non Sequitur} visualizations to reveal some information from the data, but they do not.
They are mostly created for decorative purposes.
On the other hand, even when the visualization is visualizing data, it can be based on \tag{Selective Data}, also known as cherry-picking\cite{klass2008just}.
The visualization selects the partial truth to tell and hides any inconvenient elements.
These pickings can be picking a specific period, specific comparison opponents, or selecting a particular set of data points.
\tag{Dubious Data} are reported cases that cast doubt on the underlying data or \leo{questionable}{} data source.
\tag{Too Few Data Points} is an issue that arises when connecting two dots with a straight line to form a line chart.

The \cat{Lie with Statistics} category is statistical transformations that manipulate the data into shapes that convey the intended messages.
Without the purpose of grouping data points, a \tag{Discretized Continuous Variable} unnecessarily becomes a categorical variable, and the values near the boundaries of the splits become prominently different.
A typical example is applying categorical rainbow colors to continuous variables
\leo{(Figure 2f).}{when sequential colors can be used instead.} 
\leo{}{This exaggerates the boundary cases.}
\leo{}{Take the color scheme in \autoref{fig:stage1issues}f as an example, when 46.11\% is red and 46.12\% is blue, that 0.01\% difference becomes very significant, which is misleading.}
The common issue of population maps is an example of \tag{Missing Normalization}.
It shows meaningless comparisons when the numbers are in absolute values instead of per capita.
\tag{Two-way Normalization} does not miss out on normalization, but arbitrarily interprets a two-way table by dividing the numbers by their column or row total \cite{cairo2019charts}.
\tag{Inappropriate Item Order} can hide patterns or show misleading trends by putting the items in an arbitrary order, like \autoref{fig:stage1issues}h.
\tag{Inappropriate Metric} is a carefully picked metric, threshold, or definition of grouping to support unconvincing claims.
\tag{Trend Line on Random Data} is a regression line on weakly correlated data that misleads people into thinking there is a trend.
\tag{Questionable Prediction} extrapolates the current trend beyond the present time without stating how the projection is made.
\tag{Inappropriate Use of Accumulation} can hide a recent declining trend by showing the accumulated historical values.
\tag{Inappropriate Aggregation Granularity} is similar to applying an inappropriate magnification level on a microscope, both of which lead to the wrong conclusion or false findings.

\subsection{Visualization Design Stage: Choices in Visual Encoding}


\begin{figure*}
  \centering
  \includegraphics[width=\linewidth, height=\textheight, keepaspectratio, clip, trim=0cm 0cm 0cm 0cm]{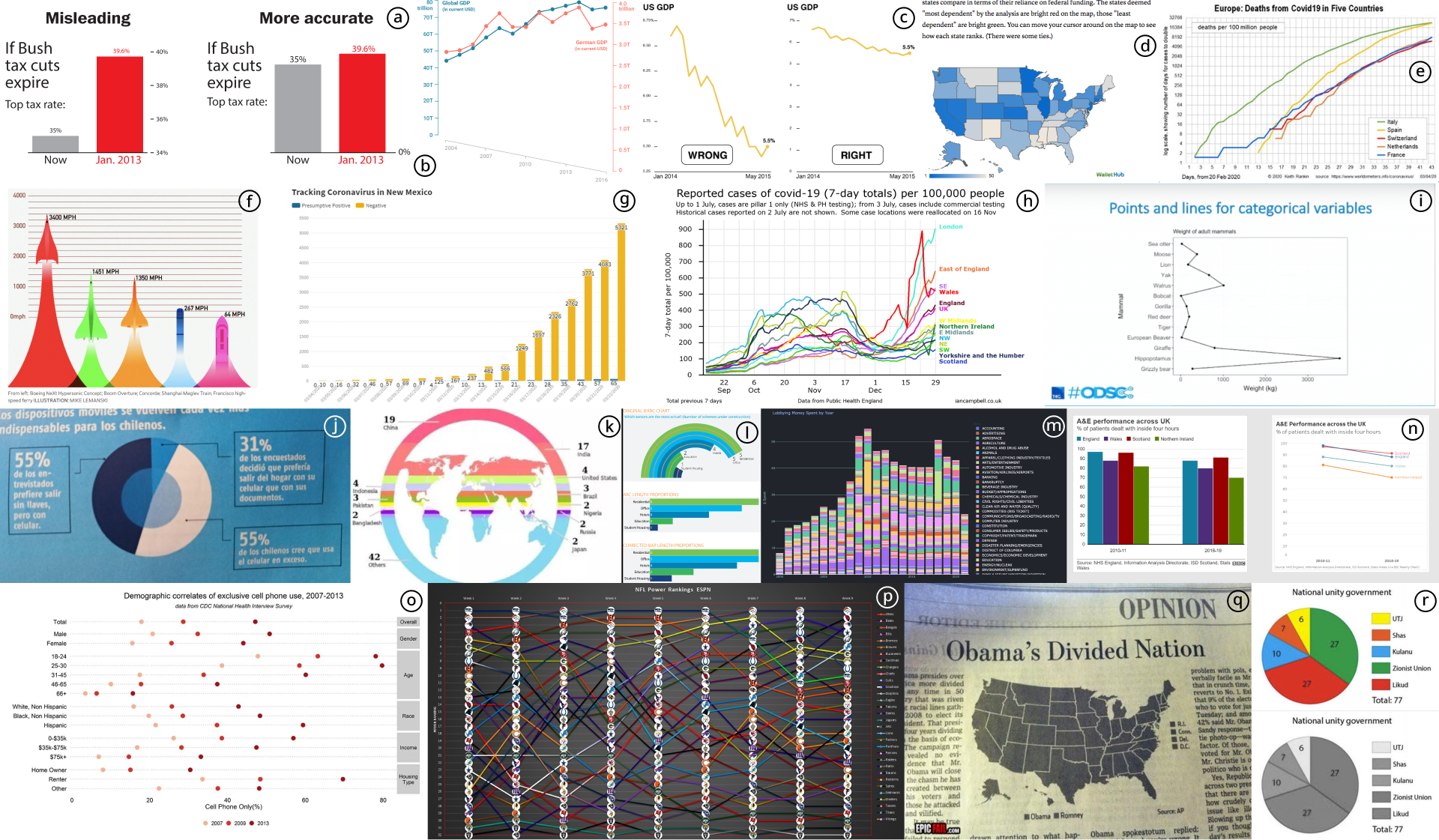}
  \caption{Issues in the Visualization Design Stage. These issues are (a) Truncated Axis, (b) Dual Axis, (c) Inappropriate Axis Range, (d) Inverted Axis, (e) Log Scale, (f) Extended Axis, (g) Data of Different Magnitudes, (h) Linear Scale on Exponential Data, (i) Inappropriate Use of Line Chart, (j) Inappropriate Use of Pie Chart, (k) Confusing Chart Type, (l) Misusing Circular Layout, (m) Inappropriate Use of Stacked, (n) Inappropriate Use of Bar Chart, (o) Inappropriate Use of Scatterplot, (p) Overusing Colors, (q) Indistinguishable Colors, and (r) Color Blind Unfriendly. See the Appendix for an explanation of each visualization, accompanied by a high-resolution version of the image.}
  \label{fig:stage2issues}
  \squeezeafterfigure
\end{figure*}

\cat{Choice of Axis} is the most common visualization design decision.
While visualizing two series with different magnitudes in one chart, a decision is made on the choice of the axis.
\tag{Dual Axis} (\autoref{fig:stage2issues}b) and \tag{data of different Magnitudes} (\autoref{fig:stage2issues}g) have the same underlying problem: two series with different magnitudes.
The series with a larger magnitude will dominate the one with a smaller magnitude, and the reader can hardly see its changes.
While plotting the two series with two different axes, the chart becomes misleading if the reader is unaware.
Even when we fix the problem to only one axis, choosing an appropriate axis range is not easy \cite{witt2019graph, correll2020truncating}.
Choosing an \tag{Inappropriate Axis Range} (\autoref{fig:stage2issues}c) has the same over-exaggerated or understated result as \tag{Truncated/Extended Axis} (\autoref{fig:stage2issues}a and \autoref{fig:stage2issues}f).
An \tag{Inverted Axis} is a decision on the axis direction.
Conventionally, the axis develops from smaller numbers, moving upward to larger numbers, but this is not the case when the number represents a ranking.
Readers expect smaller numbers at the top instead of at the bottom.
The same principle applies to color saturation and brightness.
\autoref{fig:stage2issues}d is a choropleth map with an inverted color scheme on ranking data.
The solid blue represents the least dependent states, and grey--a less saturated color--represents the most dependent states while the readers expect the opposite.
\tag{Log Scale} and \tag{Linear Scale on Exponential Data} are two sides of the same coin, so when should you apply log scale and when should you not?
Again, this relates to the convention, visualization literacy, and the readers' expectation.
Plotting exponential growth data on a linear scale shows fewer meaningful patterns, but a log scale can be misleading if the readers focus only on the curve.

The common cause of inappropriate \cat{Choice of Chart} is applying the wrong chart type to unsuitable data.
Putting categorical data on axes or swapping the x and y-axis are \tag{Inappropriate Use of Line Chart} (\autoref{fig:stage2issues}i).
Plotting a pie chart on data that has no part-to-whole relationship is the most common cause of \tag{Inappropriate Use of Pie Chart}.
\tag{Inappropriate Use of Stacked} is when too many layers are stacked, making it incomprehensible for the reader.
\tag{Misusing Circular Layout} attempts to use curved bars in a comparison or render length-encoded lines in curves, resulting in the readers getting an inaccurate comparison.
A chart has a \tag{Confusing Chart Type} because it partly belongs to a common chart but also deviates from it.
These charts are hard to decode or understand and leave readers in confusion.
Similarly, readers cannot comprehend the \tag{Inappropriate Use of Bar Chart and Scatterplots}.


When colors in the visualization are indistinguishable, it is a \cat{Color Mess}, and the reader cannot get information from the visualizations.
\tag{Overusing Colors} in a visualization can only overwhelm the reader, and the visualization becomes hideous to look at.
\tag{Indistinguishable Colors}, on the other hand, are when the same or very similar colors are used for different categories, causing the readers to mistake one category for another.
For color blind people, a \tag{Color Blind Unfriendly} visualization is similar to indistinguishable colors.

\subsection{Plotting Stage: Drawing on the Canvas}


\begin{figure*}
  \centering
  \includegraphics[width=\linewidth, height=\textheight, keepaspectratio, clip, trim=0cm 0cm 0cm 0cm]{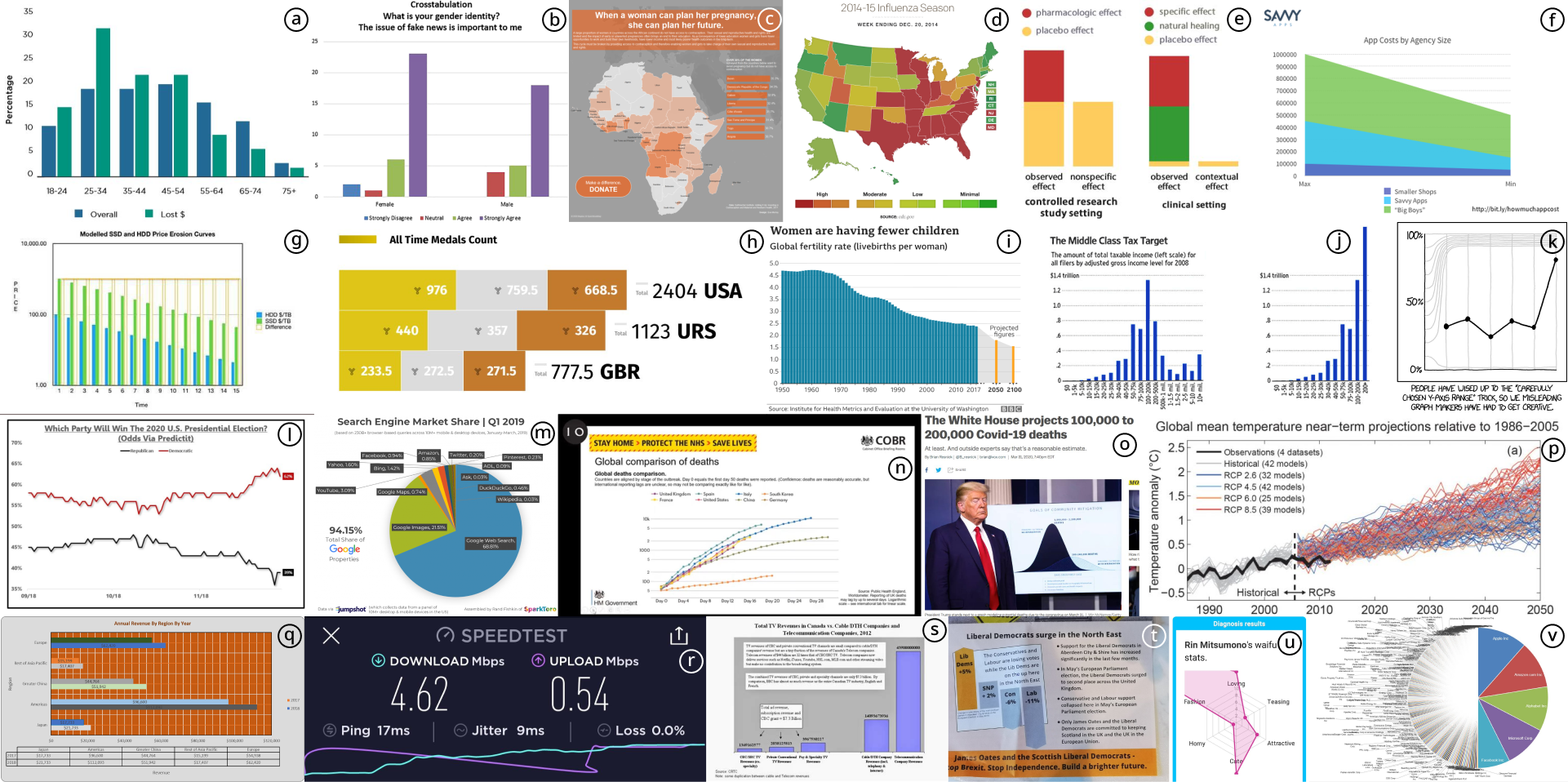}
  \caption{Issues in the Plotting Stage. These issues are (a) Missing Title, (b) Missing Axis Title, (c) Missing Legend, (d) Missing Value Labels, (e) Missing Axis, (f) Missing Axis Ticks, (g) Missing Units, (h) Misrepresentation, (i) Inconsistent Tick Intervals, (j) Inconsistent Binning Size, (k) Changing Scale, (l) Violating Color Convention, (m) Inconsistent Grouping, (n) Inconsistent Tick Labels, (o) Inconsistent Value Labels, (p) Cluttering, (q) Confusing Legend, (r) Plotting Error, (s) Missing Abbreviation, (t) Misalignment, (u) Plotting Out of Chart, (v) Illegible Text. See the Appendix for an explanation of each visualization, accompanied by a high-resolution version of the image.}
  \label{fig:stage3issues}
  \squeezeafterfigure
\end{figure*}

A chart may still be readable with minor chart elements missing, but an \cat{Incomplete Chart} is incomprehensible when missing one or more elements that are crucial to the understanding of the meaning of the chart.
A visualization with a \tag{Missing Title} keeps the reader wondering what the visualization is about and what message it is intended to convey.
\tag{Missing Axis Title} makes the meaning of the axis ambiguous.
\tag{Missing Axis} and \tag{Missing Legend} can be compensated for with labels on the visual marks.
But, when all of these elements are missing, readers cannot get any information from the visualization (\autoref{fig:stage3issues}d).

\cat{Inconsistency} occurs in arbitrarily drawn charts that break our conventional understanding of the concepts.
Readers expect the charts to be drawn according to our shared understanding of the world.
For example, drawing to scale, following convention, and using the same units across different values.
\tag{Misrepresentation} is when data values are drawn disproportionately or not to scale.
It is commonly spotted as value labels that fail to match the visually encoded geometric objects.
We expect the tick intervals in a chart to be consistent, but charts with \tag{Inconsistent Tick Intervals} violate this expectation and show false patterns.
Malicious visualization authors can use this trick to create a distorted visual perception.
\tag{Inconsistent Binning Size} means varying the boundaries of the binning groups to put more or fewer items into specific groups, resulting in a manipulated statistics calculation.
\tag{Changing Scale} means the scale changes midway.
It can be treated as a special case of inconsistent tick intervals.
The tick intervals change in the later part of the axis with or without any visual cues.
\tag{Violating Color Convention} is a color mismatch between colors and the category they represent.
\tag{Inconsistent Tick Labels} are tick labels that are not in the same format.
\tag{Inconsistent Value Labels} are the labels that are inconsistently annotated.
\tag{Inconsistent Grouping} is when some entities are grouped while others are not.
It makes the dominating entity look less dominant by splitting it into smaller members while the other groups remain grouped.


Visualizations with a \cat{Chaotic Canvas} are unreadable.
The rendered visualization is glitchy or messed up.
\tag{Cluttering} is a scalability issue related to too many data points or series clumping together.
A chart with a \tag{Confusing Legend} is difficult to comprehend its encoding or incomprehensible at all.
A \tag{Plotting Error} is a glitch in a chart as a result of software bugs or other reasons.
\tag{Plotting Out Of Chart} means plotting a data value that goes beyond the axis range, therefore, its geometric mark is out of the chart area.
\tag{Misalignment} of the items or labels results in poor readability, making it hard to perform a comparison.
\tag{Missing Abbreviation} makes the text unnecessarily long, and \tag{Illegible Text} is the text that overlaps itself or other text and becomes unreadable.

\subsection{Perception Stage: Visually Perceiving the Visualization}

\begin{figure*}
  \centering
  \includegraphics[width=\linewidth, height=\textheight, keepaspectratio, clip, trim=0cm 0cm 0cm 0cm]{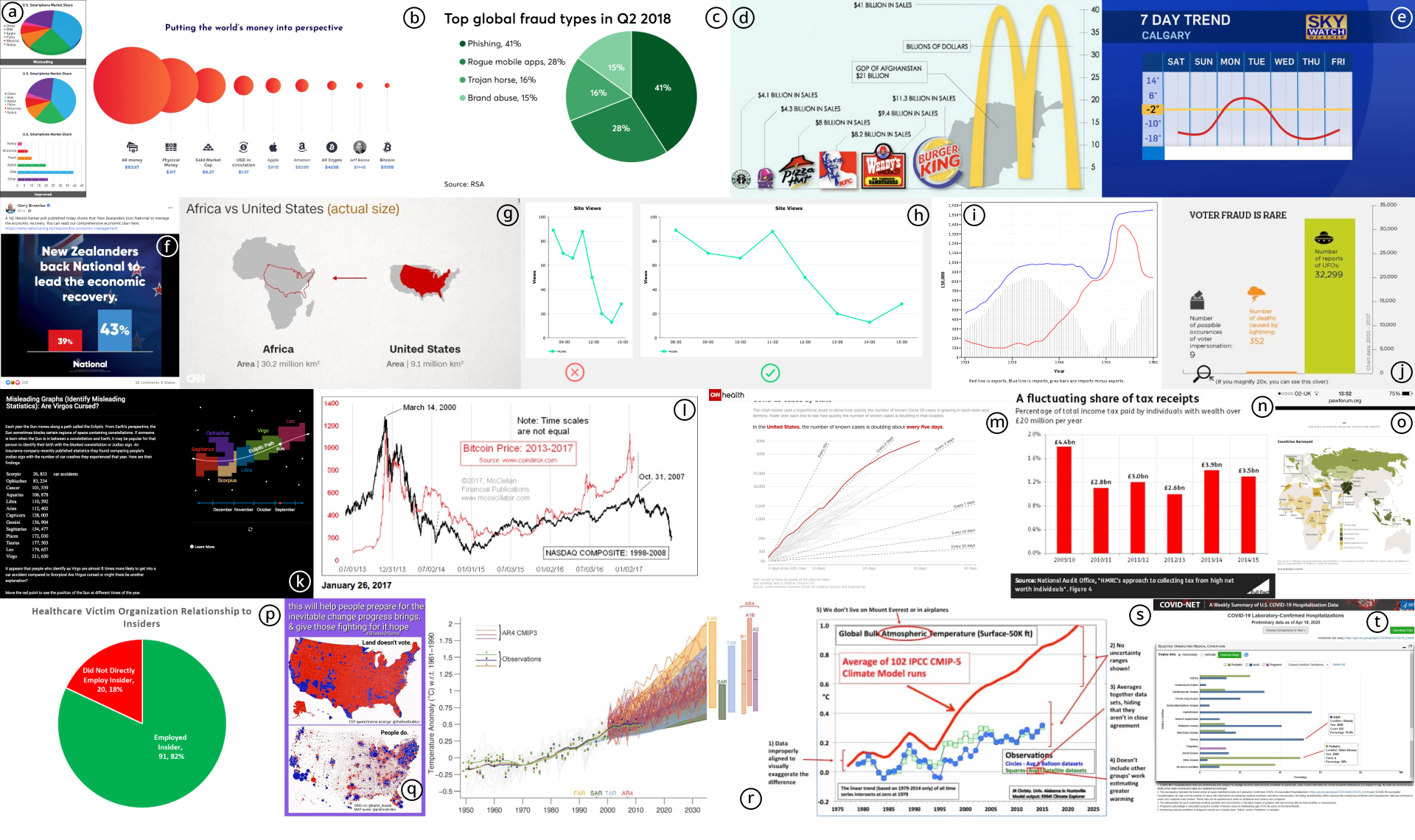}
  \caption{Issues in the Perception and Interpretation Stage. These issues are (a) 3D, (b) Area Encoding, (c) Ineffective Color Scheme, (d) Pictorial Area Encoding, (e) Inappropriate Use of Smoothing, (f) Distractive Value Labels, (g) Map Projection Distortion, (h) Inappropriate Aspect Ratio, (i) Sine Illusion, (j) Invalid Comparison, (k) Correlation Not Causation, (l) Pattern Seeking, (m) Misleading Claim, (n) Misleading Annotation, (o) Misleading Title, (p) Misleading Value Labels, (q) Hidden Distribution, (r) Overplotting, (s) Hidden Uncertainty, (t) Hidden Population Size. See the Appendix for an explanation of each visualization, accompanied by a high-resolution version of the image.}
  \label{fig:stage45issues}
  \squeezeafterfigure
\end{figure*}


The only category in this stage is \cat{Visual Illusion}.
The visualization is drawn to scale, but we perceive it differently.
Unjustified \tag{3D} is a perspective distortion technique.
The closer it is, the larger it looks, despite being the same size in the 3D perspective.
The closer objects are perceived as more prominent, although they may be the same or smaller.
It allows more ink for smaller objects in 3D bar charts.
\tag{Area Encoding} and \tag{Pictorial Area Encoding} are not linear encodings.
According to Stevens' power law, the exponent for area is \char`\~0.7, which means linearly encoding data values as areas leads the readers to consistently underestimate the values \cite{stevens1975psychophysics}.
\tag{Ineffective Color Scheme} are rainbow colors, categorical colors on sequential data, and sequential colors on categorical data.
Most are related to colors used on a choropleth.
Dichotomy colors can hide the underlying distribution (\autoref{fig:stage45issues}q).
A light red (50.1\%) can conceal all the blue (49.9\%), which is most commonly seen in choropleth maps concerning political voting.
\leo{Discretize continuous variables to use classed colors when sequential colors can be used instead.}{} 
\leo{This exaggerates the boundary cases.}{}
\leo{Take the color scheme in Figure 2f as an example, when 46.11\% is red and 46.12\% is blue, that 0.01\% difference becomes very significant.}{}
The commonly used Mercator projection suffers from \tag{Map Projection Distortion}, and comparing the area is inaccurate on these 2D maps.
However, we did not find any deliberately deceptive cases.
\tag{Distractive Value Labels} are the value labels that affect perceived data size.
The size of value labels provides visual cues that even dominate the data-encoded geometric objects.
\tag{Inappropriate Use of Smoothing} is the interpolation between data points.
It creates non-existing data points.
\tag{Sine Illusion} is one of the 143 optical illusions \cite{bach2006optical}.
There may be a lot of these going unspotted because most people are not aware of them.
They are hard to detect by the human eyes but are perhaps easier for machines to spot.
We came across a blog post in the coding process that discusses why the middle series of stacked area charts are consistently underestimated~\cite{zeigen2010lying}.
Both line chart and stacked area chart are suspected of this illusion.
An \tag{Inappropriate Aspect Ratio} is commonly seen in line charts, but it can also be applied to other chart types \cite{ceja2020truth}.

\subsection{Interpretation Stage: Comprehending the Message}

When visualization is being used for communication, we need to remain \leo{skeptical mind}{skeptical} about its message.
It can put two or more things together and suggest a \cat{False Linkage}.
It can also use the \cat{Power of Words} to mislead the interpretation of the visualization.
Or, you may interpret something differently if it keeps any crucial \cat{Hidden Information} from you.


\cat{False Linkage} is putting unrelated or incomparable items together.
\tag{Invalid Comparison} is like comparing apples to oranges by linking unrelated items by comparison or incomparable units/normalization.
It is used as a red herring, like comparing the chances of voter fraud (\autoref{fig:stage45issues}j).
\tag{Correlation Not Causation} is trying to lead the reader to draw a causality conclusion by falsely linking events to explain the data and give a cause to the observed data.
\tag{Pattern Seeking} is to seek a correlation from historical data to try to predict what happened in the past will happen in the future just because the pattern matches.

\cat{Power of Words} is when the text description on the chart does not match the message conveyed by the chart.
The chart may be faithfully plotted, but the text is misleading.
\tag{Misleading Claim} is the use of a chart to make a misleading claim by their interpretation through text description.
\tag{Misleading Annotation} is when the annotation in the chart is misleading.
\tag{Misleading Title} is when the title does not match the message in the chart.
\autoref{fig:stage45issues}p has \tag{Misleading Value Labels} that the interpretation at first glance is a pie chart with sectors 91.82\% and 20.18\%.

\cat{Hidden Information} is when some critical information that is not being shown on the chart leads to the drawing of incorrect conclusions or being used to support a misleading claim.
\tag{Hidden Distribution} is an underlying distribution that is not shown in the visualization, causing a misconception of the actual picture.
Charts with \tag{Overplotting} issues have visual marks in the same position.
The marks that are being overlapped are hidden from the readers, resulting in understated areas (\autoref{fig:stage45issues}r).
\tag{Hidden Uncertainty} is the case that uncertainty is not being visually represented when it matters.
\tag{Hidden Population Size} is when the population size matters to the statistical summary but is not shown.

\section{What Have We Learned From Misinformative Visualizations?}
\label{discussion}

Throughout the process of coding and forming the taxonomy, we found four promising future research directions to help better understand misleading visualizations.


\textbf{Understanding the designers’ dilemma.}
In some misleading cases, we found that the visualization authors have already tried their best not to mislead the reader by providing visual cues to mitigate any misleading effect.
Yet, the resulting visualization is still misleading.
Given a data series mixed with dense historical data and sparse projected data, despite the best efforts of the author to avoid misleading the readers by providing visual cues, the chart with inconsistent tick intervals in \autoref{fig:stage3issues}i still misleads readers.
A similar difficulty arises when the binning or intervals, like age or income groups, are unevenly set during the data collection process (\autoref{fig:stage3issues}j).
The case of dual axis has the root cause of visualizing two data series of different magnitudes in one chart.
The smaller magnitude series is dominated when both series are put under a unified axis scale (\autoref{fig:stage2issues}g).
The discussions of these issues are valuable guidance for practitioners \cite{ellis2016dual, muth_2018}.
Besides the problems inherited from the data, the limitations may come from the presentation medium.
Projecting the spherical Earth onto a 2D world map means distortion is unavoidable (\autoref{fig:stage45issues}g).
Finding and understanding these dilemmas in visualization designs is an important step toward developing practical design guidelines.
The great design of railway maps is possible only when the designers realize the constraints they are facing \cite{venetikidis_2012}.

\textbf{The role of conventions and data literacy education.}
Breaking convention is a major source of misperception \cite{correll2017black}.
The visualizations we encountered during the coding process reminded us of the convention changes from culture to culture and time to time.
A good example of such a case is the use of log scales in charts concerning COVID-19 cases.
At the beginning of the outbreak, people found the log scale misleading because it flattened the curve.
However, to people who had learned to read log scale charts, not using log scale is also misleading due to its exponential nature (\autoref{fig:stage2issues}h).
The discussion of breaking conventions is always fuzzy because we do not have an agreed set of conventions \cite{chen2020isms}.
It becomes even trickier when considering its shifting nature across cultures and time.
Data literacy education plays a vital role in shaping our expectations on how to read visualizations.
Like the log scale example, people who learned to read log scale charts expect exponential data to be plotted in a log scale, while others find log scale misleading rather than informative.
Clarifying the existing conventions and their variations across cultures is essential to progressing our discussion on broken conventions.
The same importance lies in shaping the conventions through data literacy education.

\textbf{The unexplored space of informal fallacies in visualization.}
Data visualizations are seldom standalone or self-explanatory.
They are often used in the context of arguments, and informal fallacies in arguments also apply to visualizations.
It happens in the input and interpretation stages.
Invalid input is fated to result in a false visualization.
Selectively choosing the most favorable data to visualize falls into the cherry-picking fallacy (\autoref{fig:stage1issues}b).
Even with valid data, \autoref{fig:stage1issues}i shows how to create a straw man by carefully crafting a metric to make a visualization supporting a misleading claim.
When interpreting a voting result visualization, misinterpreting the territorial area as population creates an illusion of a landslide victory (\autoref{fig:stage45issues}q), which falls into red herring fallacy.
Other fallacies are the false analogy fallacy (as \tag{invalid comparison} in \autoref{fig:stage45issues}j), post-hoc fallacy (as \tag{correlation not causation} in \autoref{fig:stage45issues}k), and slippery slope fallacy (as \tag{hidden uncertainty} in \autoref{fig:stage45issues}s).
Informal fallacies in visualizations have very different representations and are hard to spot, so we suspect more--if not all--of the informal fallacies are applicable to visualizations.


\textbf{Does there \leo{exists}{exist} a type of chart that is immune to misleading the audience?}
There are chart types that do not apply to certain types of issues.
For example, axis-related issues do not apply to pie charts because there are no axes in pie charts.
We wonder if a chart type is immune to misleading tricks.
Conversely, can we always make a chart misleading?
From our collected samples, we have only found two cases of uninformative word clouds (\tag{non sequitur} and \tag{missing legend}).
Still, we did not find any reported instances of misleading word clouds.
Word cloud seems to be a candidate that is immune to misleading the audiences.
Unfortunately, a previous synthetic study has shown that word clouds can be injected with manipulated data without it being spotted \cite{wickham2010graphical}.
We are afraid that no chart is immune to misleading the audience.
Instead, some are stealthier and are rarely spotted, which is even more dangerous.
This brought our attention to similar cases of sine illusion and distorted map projections.
Both cases are more prevalent than expected as they were not pointed out and discussed as frequently as others.
Finally, instead of designing a type of chart that is immune to misleading the audience, we must rely on the authoring process to better prevent misleading elements from sneaking into the visualizations.
That is, fortifying each stage of the visualization pipeline, validating and verifying the produced visualizations.
The preliminary work on visualization linter \cite{mcnutt2018linting, zhengvisualization, mcnutt2020surfacing, hopkins2020visualint, chen2021vizlinter} is a promising direction, and this work completes the missing link between the linters and real-world examples \cite{hopkins2020visualint}.

\section{Conclusion}
\label{conclusion}

Visualization has persuasive power, and this power can be used to guide people in the wrong direction.
This work aims to enrich the discussion on misleading visualizations.
From the collected dataset and the taxonomy, we found a large issue space that leads to the resulting uninformative or misleading visualization.
The ever-increasing use of visualization in public discourse as part of arguments also falls into informal fallacies.
On one side, educating the general public on better data and visualization literacy is an urgent matter that is also welcomed by students \cite{bergstrom2017calling}.
On the other side, expanding the discussion scope will help us learn more about the nature of misleading visualizations.
We will continue our research in this area, opening up the aforementioned research directions and developing preventative measures for countering misleading visualizations.
\bibliographystyle{eg-alpha-doi}
\bibliography{egbibsample.bib}



\end{document}